\documentclass{article}%
\usepackage{amssymb}
\usepackage{amsmath}
\usepackage{amsfonts}
\usepackage{graphicx}
\begin{document}

\title{Superposition of fields of two rotating charged\\ masses in General Relativity and existence\\ of equilibrium configurations
}
\author{G.A. Alekseev$^{\text{*)}}$ and V. A. Belinski$^{\text{**)}}$\\$^{\text{*)}}$Steklov Mathematical Institute, Gubkina 8, Moscow 119991,\\Moscow, Russia,\textit{ G.A.Alekseev@mi-ras.ru}\\$^{\text{**)}}$ICRANet, Piazzale della Repubblica, 10, 65122 Pescara, Italy;\\Rome University "La Sapienza", 00185 Rome, Italy;\\IHES, F-91440 Bures-sur-Yvette, France,\textit{ belinski@icra.it }}
\date{}
\maketitle

\begin{abstract}
It is known that two Reissner-Nordstrom black holes or two overextreme
Reissner-Nordstrom sources cannot be in physical equilibrium. In the static case such equilibrium is possible only if one of the sources is a black hole and another one is a naked singularity. We define the notion of physical
equilibrium in general (stationary) case when both components of a binary system are rotating and show that such system containing a Kerr-Newman
black hole and a Kerr-Newman naked singularity also can stay in physical
equilibrium. The similar question about the system of two charged rotating black holes or two rotating overextreme charged sources still remaines open.
\end{abstract}


\section{Introduction}

In the non-relativistic physics two particles can be in equilibrium if the
product of their masses is equal to the product of their charges (we use the
units $G=c=1$). However, the question on the existence of an analogue of such
equilibrium state in General Relativity is far from being trivial. Besides the
natural mathematical complications, in General Relativity there arise two different types of the "point" centers, namely Kerr - Newman black holes and Kerr - Newman naked singularities. Therefore, seeking equilibrium configurations, one has to consider all three possible types of binary systems
\[\begin{array}{l}
\bullet\hskip1ex\text{black hole - black hole}\\
\bullet\hskip1ex\text{black hole - naked singularity}\\
\bullet\hskip1ex\text{naked singularity - naked singularity}
\end{array}
\]
and analyse in each case all  physical and geometrical conditions which are necessary for equilibrium. In particular, in each case a physically defined  distance between these objects should exist in the corresponding solution.

When the Inverse Scattering Method (ISM) has been adopted for integration of
the Einstein and Einstein-Maxwell equations, it was shown that Kerr-Newman black holes and Kerr-Newman naked singularities represent nothing
else but stationary axially symmetric solitons. Then by the ISM machinery one
can obtain the families of exact stationary axially symmetric
solutions of these equations containing any number of such solitons centralized at different points of the symmetry axis. The mathematical construction of such solutions do not represents any principal difficulties apart from the routine calculations in the framework of the well developed soliton generating procedure which allows to insert any number of solitons into a given background space-time. However, it is quite intricate task to single out from these families the \textit{physically} reasonable cases which correspond to a real equilibrium states of charged rotating black holes and naked singularities interacting with each other because in general, formally constructed stationary axisymmetric soliton solutions possess some features unacceptable from the physical point of view. These unwanted traits are due to the presence in the solutions of exotic peculiarities such as:
\begin{itemize}
\item{non-vanishing global NUT parameter,}
\item{closed time-like curves around those parts of the symmetry axis which are outside the sources,}
\item{angle deficit (or excess) at the points of the symmetry axis,}
\item{non-zero physical magnetic charges of the sources,}
\item{existence of some additional singularities on or off the axis}
\end{itemize}
The global NUT parameter is incompatible with asymptotic flatness of the
space-time at spatial infinity. Keeping in mind the physical applications, we should exclude also the appearance outside the sources of closed timelike curves which may arise in formally constructed solutions. Besides that, the angle deficit or excess at the points of the symmetry axis give rise to the well known conical singularities violating the local Euclidness of space at these points (it can be treated as some singular external strut or string preventing the sources to fall onto or to run away from each other). Also the magnetic charges of each of the sources should be excluded since their existence contradicts to the present physical experience.\footnote{In our pesent consideration we use the condition that the total magnetic charge as well as the magnetic charges of each of the sources are absent. However, if one is interested in consideration of the equilibrium configurations with magnetic charges also, it is easy to exclude from our equilibrium conditions the corresponding conditions for their absence, because such magnetic charges in the solution (\ref{ErnstPotentials}) -- (\ref{c0}) are present initially.}
All five aforementioned phenomena have nothing to do with a real equilibrium of the physical bodies and the corresponding equilibrium solution should be free of such pathologies. To single out from the families of soliton solutions the solutions without all listed above physically unacceptable properties, one needs to impose on the parameters of these families of solutions some constraints  which lead to a system of algebraic equations. The problem is that these equations, even for the simplest case of two objects, are extremely complicated and it is difficult to resolve them in an exact analytical form in order to show directly whether they have physically appropriate solutions compatible with the condition of existence of a positive distance between the sources.

The aforementioned nuisances constitute the real troubles only in the general
case of rotating sources. The static case is more simple and it would
be not an overstatement to say that for the case of two non-rotating charged
objects the problem have been solved completely. The first indications that
two static charged masses can stay in real physical equilibrium without any
struts between them and without any other pathologies came from the results of
Bonnor \cite{Bon} and Perry and Cooperstock \cite{Per}. In \cite{Bon} it was
analyzed the static equilibrium condition for a charged test particle in the
Reissner-Nordstrom field and it was shown that such test body can be at rest
in the field of the Reissner-Nordstrom source only if they both are either
extreme (the charge equal to mass) or one of them is of a black hole type (the charge is less than the mass) and the other is of naked singularity type (the charge is grater than the mass). There is no way for equilibrium in cases when both masses are either of naked singularity type or both are of black hole  type. The more solid arguments in favour of existence of a static equilibrium configuration for the black hole  - naked singularity system was presented in \cite{Per}, where both sources have been treated exactly, that is no one of the components was considered as test particle. These results have been obtained there by numerical calculations and three examples of numerical solutions of the equilibrium equation have been demonstrated. These solutions can correspond to the equilibrium configurations free of struts, though the authors have not been able to show the existence of a positive definite distance between the sources. The authors of \cite{Per} also reported that a number of numerical experiments for two black holes and for two naked singularities showed the negative outcomes, i.e. all tested sets
of the parameters was not in power to satisfy the equilibrium equation. These
findings were in full agreement with Bonnor's test particle analysis.

The explicit solution of the problem in static case have been
presented in paper \cite{Alekseev-Belinski:2007} where it was constructed the exact analytic static solution of the Einstein-Maxwell equations for two charged massive sources separated by the well defined positive distance and free of struts or of any other unphysical properties \footnote{More details of derivation of this solution an interested reader can find in the paper \cite{AB2}.}. We showed also that such solution indeed exists only for the black hole -- naked singularity system and it is impossible to have the similar static equilibrium state for the pairs black hole -- black hole or naked singularity -- naked singularity. After these results the natural question arises whether the analogous physical equilibrium exists for two \textit{rotating} sources. It turns out that the answer is affirmative and in the present paper we demonstrate the \emph{six-parametric} exact equilibrium solution of the Einstein-Maxwell equations for two rotating charged objects one of which is a black hole and another one is a naked singularity.

It is worth mentioning that in the literature one can find many formal
mathematical constructions of particular cases of doble-soliton stationary solutions. However, all of them have one and the same weakness: they cannot be accepted as physical equilibrium states because each of them contains at least one of the five pathological traits listed above. The present article proves the existence of the physical equilibrium states of two rotating sources of the black hole -- naked singularity type which is free of all these pathologies and depends on \emph{six} independent parameters. We restrict ourselves
namely by the system of rotating bodies of the black hole - naked singularity type, because it is known that for such system the physical equilibrium in statc case indeed is possible. As we said already this is only the case for the static physical equilibrium configuration and it is natural to expect that it can be extended also for the rotating sources of these types. Nevertheless, it is necessary to emphasize that in spite of the known non-existence of physical equilibrium for the systems black hole - black hole and naked singularity - naked singularity in static case, we cannot assert the non-existence of physical equilibrium for such systems also for rotating configurations. May be the rotations can create the additional forces which can overcome the "no go" result for the static case for the systems black hole - black hole and naked singularity - naked singularity and permit them to stay in the physical equilibrium (although it seems to be a little bit strange, because such equilibrium would have no limit to the case of vanishing rotations). Thus, for binary systems black hole - black hole and naked singularity - naked singularity, the question on the possibility of physical equilibrium at present remaines open.

Besides that, we had to clarify also an important question concerning the absence in our solution of any other singularities anywhere outside two Kerr-Newman sources. At the last section of the manuscript, we construct an asymptotic representation of our solution in the whole space-time region outside two Kerr - Newman sources -- the black hole and naked singularity for the case of rather large coordinate distance separating these two sources. The regular character of this asymptotic representation indicates the non-existence in this solution of other singularities which could be considered as supplement sources of gravitational and electromagnetic fields in our solution.

\section{General properties of soliton solutions}

In the context of the Inverse Scattering Method (ISM), the gravitational solitons  as exact solutions of pure gravity Einstein equations have been introduced in the papers \cite{BZ1, BZ2}. The generalization of this technique for the coupled gravitational and electromagnetic fields was constructed in \cite{A1}. Its more detailed description can be found in \cite{A2} and in the book \cite{BV}. In this generalized approach one starts from some given background solution of the Einstein-Maxwell equations and generates on this background any desired number of solitons. We have to do here with the linear spectral differential equations (Lax pair) for the $3\times3$ matrix function $\Psi(\rho,z,w)$, where $w$ is a complex spectral parameter independent of coordinates $\rho,z$. First of all, for chosen background solution of the Einstein-Maxwell equations we have to find from the Lax pair the corresponding background spectral matrix $\Psi_{0}(\rho,z,w)$. Using the ISM dressing procedure it is possible to find explicitly the spectral matrix $\Psi_{n}(\rho,z,w),$ corresponding to the new solution containing $n$ solitons generating on the background space-time and   extract then the new metric and new electromagnetic potential from this $\Psi
_{n}.$ The solitonic field added to the background can be characterized by the
matrix $\Psi_{n}\Psi_{0}^{-1}-I$ which is a meromorphic (with respect to the
spectral parameter $w$) matrix function tending to zero in the limit
$w\rightarrow\infty$ and having $n$ simple poles in the complex plane of the
parameter $w$ (one pole for each soliton).

In pure gravity case some of these poles can be located at the real axis of
the $w$-plane and the corresponding sources have horizons (that is they are of
the black hole type) while complex poles generate objects with naked singularities.
However, in the presence of electromagnetic field the formal machinery of the
ISM developed in \cite{A1} in general does not allow poles to be located
at real axis which means that by this method one can produce solutions
containing sources without the horizons, i.e. only of the naked singularity type.\footnote{We said "in general" because
it can be shown that the ISM considered in \cite{A1} can be adjust also to the
real $w$-poles but only for that special restriction on the parameters of the
solutions which correspond to the extreme black holes.} Nevertheless, also in
this case after one obtains the final form of solution it is possible to
forget the way how it was derived and to continue the solution analytically in
the space of its parameters in order to get the complete family containing
solutions with real metric of the physical signature and with horizons as
well. However, the technical procedure how to do this is simple only for the
case of one-solitonic solution (that is for the Kerr-Newman case) and some
simple enough generalization of such procedure was found also for two static
solitonic objects \cite{AB3}. In the general case of two rotating sources (the
corresponding 12-parametric solitonic solution have been constructed in
\cite{A3}) this task is much more complicated. Fortunately, there is an
effective way to get over this difficulty. Because we need to construct
solution of the black hole - naked singularity type, we can consider the Kerr-Newman black hole as new background (instead of the flat space-time) and insert to it one soliton of naked singularity type. This is exactly what can be done easily with the generating technique proposed in \cite{A1} and what we are interested in. The exact expressions (in terms of the Ernst potentials) for the solution together with the proof that all conditions of the physical equilibrium can be satisfied are given below.

\section{Superposition of fields of the Kerr - Newman black hole and naked singularity}

For stationary axisymmetric soliton solutions of Einstein - Maxwell equations, metric and electromagnetic potential components in Weyl coordinates $\left(  t,\rho,z,\varphi\right)$ take the forms:
\begin{equation}\label{FieldComponents}
\begin{array}{c}
ds^{2}=-f\left(  d\rho^{2}+dz^{2}\right)  +g_{tt}dt^{2}+2g_{t\varphi
}dtd\varphi+g_{\varphi\varphi}d\varphi^{2},\\[1ex]
g_{tt}g_{\varphi\varphi}-g_{t\varphi}^{2}=-\rho^{2}, \\[1ex]
A_{t}=A_{t}(\rho,z),\text{ \ }A_{\varphi}=A_{\varphi}(\rho,z),\text{
\ }A_{\rho}=0,\text{ \ }A_{z}=0\text{ },
\end{array}
\end{equation}
where all metric coefficients depend only on the variables $\rho,z$.
The Lorentz signature of this metric implies that the conformal factor $f>0$.

Our solution depends on eight real and two complex constant parameters
\begin{equation}\label{Parameters}
\{m_{\scriptscriptstyle{0}},\,a_{\scriptscriptstyle{0}},\, b_{\scriptscriptstyle{0}},\,e_{\scriptscriptstyle{0}}\},\quad
\{m_{s},\,a_{s},\,b_{s},\,e_{s}\},\quad l=z_{2}-z_{1}>0\quad\text{and}\quad
c_{\scriptscriptstyle{0}}\text{ },%
\end{equation}
where the parameters $e_{\scriptscriptstyle{0}}$ and $e_{s}$ are complex and the others are real. The parameters with the suffix \textquotedblleft$0$\textquotedblright%
\ are related to the background solution (black hole) and the parameters with
the suffix \textquotedblleft${s}$\textquotedblright\ are the parameters of a
soliton we add to the black hole background. The
parameter $l$ (which was chosen positive for definiteness) characterizes a
$z$-distance between the sources because $z_{1}$ and $z_{2}$ determine
respectively the location of a black hole and a naked singularity on the axis.
The constant $c_{0}$ is an arbitrary multiplier in front of the metric
coefficient $f$ \ in (\ref{FieldComponents}) which should be chosen in accordance, e.g.,
with the condition of regularity of the axis at spatial infinity. It is
convenient to use two functions of these parameters -- the real
$\sigma_{0}$ and  pure imaginary $\sigma_{s}$, such that
\begin{equation}\label{sigmas}
\sigma_{0}^{2}=m_{0}^{2}+b_{0}^{2}-a_{0}^{2}-e_{0} \overline{e}_0\geq
0,\qquad\sigma_{{s}}^{2}=m_{{s}}^{2}+b_{{s}}^{2}-a_{{s}}^{2}-e_{s} \overline{e}_s\leq0.
\end{equation}

Though our stationary axisymmetric solution depends on two Weyl
coordinates $\rho,z$ only, it is more convenient to express it in terms of
the so called \textquotedblleft bipolar\textquotedblright\ coordinates -- two
pairs of polar coordinate $(x_1, y_1)$  and $(x_2, y_2)$ centered respectively at the location of a black hole (the coordinates with the suffix $1$) and at the location of a naked singularity (the coordinates with the suffix $2$). Of course, these four
coordinates should satisfy two additional constraints and each of these four
coordinates can be expressed in terms of Weyl coordinates $\rho,z$. The
corresponding defining relations take the forms
\begin{equation}\label{Weyl}
\left\{\begin{array}{l}
\rho=\sqrt{x_{1}^{2}-\sigma_{0}^{2}}\sqrt{1-y_{1}^{2}}\\[0.5ex]
z=z_{1}+x_{1} y_{1}
\end{array}\right.\quad\text{and}\quad
\left\{\begin{array}{l}
\rho=\sqrt{x_{2}^{2}-\sigma_{{s}}^{2}}\sqrt{1-y_{2}^{2}}\\[0.5ex]
z=z_{1}+l+x_{2}y_{2}
\end{array}\right.
\end{equation}
It is worth to note that $z_{1}$ is not an essential parameter because
it determines a shift of the whole configuration of the sources and their  fields along the axis. The inverse relations for coordinates
$(x_{1},y_{1})$ corresponding to real $\sigma_{0}$ are
\begin{equation}\label{x1y1}
\begin{array}{l}
x_{1}=\dfrac{1}{2}\left[  \sqrt{(z-z_{1}+\sigma_{0})^{2}+\rho^{2}}%
+\sqrt{(z-z_{1}-\sigma_{0})^{2}+\rho^{2}}\right]  ,\\[1ex]%
y_{1}=\dfrac{2(z-z_1)}{\sqrt{(z-z_{1}+\sigma_{0})^{2}+\rho^{2}}
+\sqrt{(z-z_{1}-\sigma_{0})^{2}+\rho^{2}}}  .
\end{array}
\end{equation}
For the coordinates $(x_{2},y_{2})$ corresponding to imaginary $\sigma_{s}$
($\sigma_{{s}}^{2}<0$) the similar relations are more complicated
($z_{2}=l+z_{1}$):
\begin{equation}
\begin{array}{l}\label{x2y2}
x_{2}=\sqrt{\dfrac{1}{2}\left[(z-z_{2})^{2}+\rho^{2}+\sigma_{{s}}%
^{2}\right]+\dfrac{1}{2}\sqrt{\left[  (z-z_{2})^{2}+\rho^{2}+\sigma_{{s}%
}^{2}\right]  ^{2}-4\sigma_{{s}}^{2}(z-z_{2})^{2}}},\\[2ex]
y_{2}=\dfrac{z-z_2}{\sqrt{\dfrac{1}{2}\left[(z-z_{2})^{2}+\rho^{2}+\sigma_{{s}}%
^{2}\right]+\dfrac{1}{2}\sqrt{\left[  (z-z_{2})^{2}+\rho^{2}+\sigma_{{s}%
}^{2}\right]  ^{2}-4\sigma_{{s}}^{2}(z-z_{2})^{2}}}}.
\end{array}
\end{equation}
Sometimes it is convenient also to use instead of pairs of coordinates
$(x_{1},y_{1})$ and $(x_{2},y_{2})$ the pairs of quasi-spherical coordinates
$(r_{1},\theta_{1})$ and $(r_{2},\theta_{2})$:
\begin{equation}\label{r1r2}
\left\{\begin{array}{l}
x_{1}=r_{1}-m_{0},\\
y_{1}=\cos\theta_{1},
\end{array}\right.
\qquad\qquad
\left\{\begin{array}{l}
x_{2}=r_{2}-m_{{s}},\\
y_{2}=\cos\theta_{2}.
\end{array}\right.
\end{equation}
The Ernst potentials and the conformal factor $f$ for our solution are:
\begin{equation}\label{ErnstPotentials}
\mathcal{E}=1-\dfrac{2(m_{0}-ib_{0})}{\mathcal{R}_{1}}- \dfrac{2(m_{s}-ib_{s})}{\mathcal{R}_{2}%
},\qquad\Phi=\dfrac{e_{0}}{\mathcal{R}_{1}}+ \dfrac{e_{s}}{\mathcal{R}_{2}},
\end{equation}
\begin{align}
\dfrac{1}{\mathcal{R}_{1}}  &  =\dfrac{x_{2}+ia_{s}y_{2}+K_{1}(x_{2}-\sigma_{s}
y_{2})+L_{1}(x_{1}+\sigma_{0}y_{1})+S_{0}\left(  x_{2}+\sigma_{s}y_{2}\right)
}{D},\label{CalR1R2}\\
\dfrac{1}{\mathcal{R}_{2}}  &  =\dfrac{x_{1}+ia_{0}y_{1}+K_{2}(x_{1}-\sigma_{0}
y_{1})+L_{2}(x_{2}+\sigma_{s}y_{2})}{D},\nonumber
\end{align}
\begin{align}\label{Dvalue}
D  &  =(x_{1}+ia_{0}y_{1}+m_{0}-ib_{0})\left[  x_{2}+ia_{s}y_{2}+m_{s}%
-ib_{s}+S_{0}\left(  x_{2}+\sigma_{s}y_{2}\right)  \right]-\\
&  -\left[  m_{0}-ib_{0}-K_{2}(x_{1}-\sigma_{0}y_{1})-L_{2}(x_{2}+\sigma
_{s}y_{2})\right]  \times\nonumber\\
&  \times\left[  m_{s}-ib_{s}-K_{1}(x_{2}-\sigma_{s}y_{2})-L_{1}(x_{1}%
+\sigma_{0}y_{1}\right]  ,\nonumber
\end{align}
\begin{align}
K_{1}  &  =\dfrac{ia_{s}-\sigma_{s}}{\sigma_{0}+\sigma_{s}+l},
\quad L_{1}=\dfrac{(m_{0}+ib_{0})(m_{s}-ib_{s})-\overline{e}_{0} e_{s}
  }{(ia_{0}-\sigma_{0})(\sigma_{1}+\sigma_{2}%
+l)},\\
K_{2}  &  =\dfrac{ia_{0}-\sigma_{0}}{\sigma_{0}+\sigma_{s}-l},
\quad L_{2}=\dfrac{(m_{0}-ib_{0})(m_{s}+ib_{s})-e_{0}
\overline{e}_{s}}{(ia_{s}-\sigma_{s})(\sigma_{0}+\sigma_{s}%
-l)},\nonumber
\end{align}
\begin{equation}\label{S0}
S_{0}=\frac{\sigma_{0}Y_{s}\bar{Y}_{s}}{\sigma_{s}\left(  \sigma_{0}^{2}+a_{0}^{2}\right)  (ia_{s}-\sigma_{s})(l-\sigma_{0}-\sigma_{s})},
\end{equation}
\begin{equation}\label{Ys}
Y_{s}=\left(  m_{0}-i b_{0}\right)  e_{s}-\left(
m_{s}-i b_{s}\right) e_{0},
\end{equation}
\begin{equation}\label{factorf}
f=c_{0}\frac{D\bar{D}}{\left(  x_{1}^{2}-\sigma_{0}^{2}y_{1}^{2}\right)
\left(  x_{2}^{2}-\sigma_{s}^{2}y_{2}^{2}\right)  }.
\end{equation}
Here and in what follows the bar over a letter means complex conjugation.
In (\ref{c0}), $c_{0}$ is an arbitrary real constant which should be chosen as
\begin{equation}\label{c0}
c_{0}=\left\vert\vphantom{I^I_I} 1+S_{0}-(K_{1}+L_{1})(K_{2}+L_{2})\right\vert ^{-2}\text{ }
\end{equation}
in order to provide a correct limit value of the conformal factor $f$
at spatial infinity where we should have $f\to 1$.

It seems useful to recall here that in Weyl coordinates the Kerr-Newman black hole horizon corresponds to the segment on the axis of symmetry $\{\rho=0,z_1-\sigma_0\le z\le  z_1+\sigma_0\}$, while the naked singularity of the Kerr-Newman type in these coordinates is represented by a segment $\{\rho=\vert\sigma_s\vert\sin\theta,\,z=z_2,\,0\le\theta\le \pi \}$ which is orthogonal to the symmetry axis. In the Kerr-Newman naked singularity geometry this segment corresponds to a "critical" sphere $r_2=m_2$.

\section{Searches for equilibrium configurations}
The solution described above was constructed as the nonlinear superposition of gravitational and electromagnetic fields of a Kerr - Newman black hole and a Kerr - Newman naked singularity. Using the previous experience of studies of static configurations of charged massive sources in equilibrium (see \cite{Alekseev-Belinski:2007}), one may expect that in the rotating case, the equilibrium  configurations of a black hole and a naked singularity also can be expected existing. However, for arbitrary choice of (in total) twelve real parameters of our solution, this solution can not be considered as describing the physical equilibrium of these two sources in their common fields, because for such physical interpretation of the solution it is necessary that in this solution any physically pathological properties (such as closed timelike curves, conical points on the axis, the total magnetic charge and the magnetic charges of each of the sources and NUT parameters) as well as  any other sources of fields (such as additional curvature singularities on the axis or outside it) are absent.
Further in this section, we find the constraints on the parameters of our solution which provides the absence in this solution of the pathological properties mentioned above, i.e. the conditions of equilibrium of the Kerr - Newman black hole and naked singularity in their common gravitational and electromagnetic fields. The absence of any other sources of these fields in the solution (\ref{ErnstPotentials}) -- (\ref{c0}) and the solution of the equilibrium conditions will be described in the next section where we use an additional assumption that the coordinate distance $l$ separating the sources is large enough.

\paragraph{\emph{Asymptotic flatness of the solution and its physical parameters.}}
At spatial infinity, i.e. for $r=\sqrt{\rho^{2}+z^{2}}
\rightarrow\infty,$ the Ernst potentials determined by the expressions (\ref{ErnstPotentials}) -- (\ref{c0}) have the following asymptotical behaviour:
\begin{equation}\label{ErnstEF}
\begin{array}{l}
\mathcal{E}=1-\dfrac{2(M-iB)}{r}+\dfrac{(z_\ast+2 i J)y+const}{r^2}+O(\dfrac{1}{r^{3}}),\\[2ex]
\Phi=\dfrac{Q_{e}+iQ_{m}}{r}+\dfrac{(D_e+i D_m)y+const}{r^2}+O(\dfrac{1}{r^{3}}),
\end{array}
\end{equation}
where $M$ and $B$ are the total gravitational mass and total NUT parameter of
the configuration, $Q_{e}$ and $Q_{m}$ are its total electric and magnetic
charges, $J$ is the total angular momentum and $D_e$ and $D_m$ are the electric and magnetic dipole moments. The constant $z_\ast$ depends on the location of  the origin of the quasi-spherical coordinate system $(r,\theta)$ on the axis of symmetry and $y=\cos\theta$.

Direct calculations show that some of these physical parameters possess the following expressions in terms of the parameters (\ref{Parameters}) and (\ref{sigmas}):
\begin{equation}\label{Physparameters}
\begin{array}{l}
M=\text{Re}\left[  \dfrac{(m_{0}-ib_{0})(1+K_{1}+L_{1}+S_{0})+(m_{{s}}
-ib_{{s}})(1+K_{2}+L_{2})}{1+S_{0}-(K_{1}+L_{1})(K_{2}+L_{2})}\right],\\[2ex]
B=-\text{Im}\left[  \dfrac{(m_{0}-ib_{0})(1+K_{1}+L_{1}+S_{0})+(m_{{s}
}-ib_{{s}})(1+K_{2}+L_{2})}{1+S_{0}-(K_{1}+L_{1})(K_{2}+L_{2})}\right],\\[2ex]
Q_{e}=\operatorname{Re}\left[\dfrac{e_{0}
(1+K_{1}+L_{1}+S_{0})+e_{s} (1+K_{2}+L_{2})}
{1+S_{0}-(K_{1}+L_{1})(K_{2}+L_{2})}\right],\\[2ex]
Q_{m}=\operatorname{Im}\left[  \dfrac{e_{0}
(1+K_{1}+L_{1}+S_{0})+e_{s}(1+K_{2}+L_{2})}
{1+S_{0}-(K_{1}+L_{1})(K_{2}+L_{2})}\right].
\end{array}
\end{equation}
The other physical parameters of our solution, such as $J$, $D_e$, $D_m$ possess more complicate expressions and we do not present them here.
From the expressions (\ref{Physparameters}) we conclude immediately that
to have a configuration without a total NUT and
magnetic charge, we should impose on the parameters the
restrictions
\begin{equation}\label{BQm}
B=0\quad \text{and}\quad Q_{m}=0
\end{equation}
where the expressions (\ref{Physparameters}) should be taken into account.

\paragraph{\emph{Closed time-like curves.}}
If at some points of the axis (where $\rho=0$) we have $g_{t\varphi}
\neq 0$, this implies (in accordance with the relation between the metric
coefficients in Weyl coordinates $g_{tt}g_{\varphi\varphi}-g_{t\varphi}
^{2}=-\rho^{2}$) that near these points $g_{\varphi\varphi}>0$. Such
inequality means that near these points of the axis the coordinate lines of
the periodic (azimuth angle) coordinate, being closed lines, are time-like. To
avoid such trouble it is necessary to demand that on every part of the axis
outside the sources and between them $g_{t\varphi}$ should vanish. As it follows directly from the Einstein - Maxwell equations, on the axis of symmetry $\rho=0$ the value $\Omega =g_{t\varphi}/g_{tt}$ is independent of $z$ and therefore, it is constant. However, this constant can be different on different disconnected parts of the axis separated by the sources. Therefore, to exclude the existence of closed time-like curves near the axis, first of all we should impose two conditions
\begin{equation}
\Omega_{-}=\Omega_{i}=\Omega_{+}\text{ }, \label{20}%
\end{equation}
where $\Omega_{-}$, $\Omega_{i}$ and $\Omega_{+}$ are the constants which are
the values of $g_{t\varphi}/g_{tt}$ on the negative, intermediate and positive
parts of the axis respectively. If the conditions (\ref{20}) are satisfied,
the corresponding common constant value of $\Omega$ on the axis can be reduced to zero by a coordinate transformation of the form $t^{\prime}=t+a\varphi$, and
$\varphi^{\prime}=\varphi$ with an appropriate constant $a$.
In order to satisfy the conditions (\ref{20}) we calculate constants
$\Omega_{+}-\Omega_{-}$ and $\Omega_{i}-\Omega_{-}$ and put both of them to
zero. Calculations show that the first constant take simple form:
\begin{equation}
\Omega_{+}-\Omega_{-}=-4B\text{ }, \label{20-0}
\end{equation}
where $B$ is the total NUT parameter given by the formula (\ref{Physparameters}), while
the expression for the second parameter is much more complicated:
\begin{equation}\label{Omegai}
\Omega_{i}-\Omega_{-}\equiv-4B-\dfrac{\omega_{{\times}}\overline{\omega
}_{{\times}}}{(a_{{\times}}+i\sigma_{{s}})}+\dfrac{(1+2\delta)}{(1-2\delta
)}\dfrac{\mathcal{H}_{0}\overline{\mathcal{H}}_{0}}{(a_{{\times}}+i\sigma
_{{s}})\mathcal{W}_{o}}.
\end{equation}
The explicit expression for $\delta$ takes the form:
\begin{equation}\label{delta}
\delta=\dfrac{\sigma_{0}(m_{0}m_{{s}}+b_{0}b_{{s}}-q_{0}q_{s}-\mu_{0}\mu_{s}%
)}{\sigma_{0}(l^{2}-\sigma_{0}^{2}-\sigma_{{s}}^{2}-2a_{0}a_{{s}}%
)+(l\sigma_{0}+\sigma_{0}^{2}-ia_{0}\sigma_{{s}})(l-\sigma_{0}-\sigma_{{s}%
})S_{0}},
\end{equation}
where $S_{0}$ was defined in  (\ref{S0}) and
\begin{equation}
\omega_{{\times}}=m_{{\times}}-ib_{{\times}}+i(a_{{\times}}+i\sigma_{{s}%
}),\qquad\mathcal{W}_{o}=(l^{2}-\sigma_{0}^{2}-\sigma_{{s}}^{2})^{2}%
-4\sigma_{0}^{2}\sigma_{{s}}^{2}\text{ }, \label{27-1}%
\end{equation}%
\begin{equation}
\begin{array}{l}
\mathcal{H}_{0}=-2i(l-\sigma_{{s}}-ia_{0}+m_{0}+ib_{0})\overline{X}_{{\times}
}\\
\phantom{\mathcal{H}_{0}=}+(a_{{\times}}+i\sigma_{{s}}-im_{{\times}}+b_{{\times}})[(l-\sigma_{{s}}%
)^{2}-\sigma_{0}^{2}]+\\
\phantom{\mathcal{H}_{0}=}+2(a_{{\times}}+i\sigma_{{s}})\left[  (m_{0}+ib_{0})(l-\sigma_{{s}}
+ia_{0})+\sigma_{0}^{2}+a_{0}^{2}\right],
\end{array}
\end{equation}
\begin{equation}\label{Xtimes}
X_{{\times}}=(m_{0}+ib_{0})(m_{{\times}}-ib_{{\times}})-\overline{e}_{0} e_{\times}.
\qquad\qquad\qquad\qquad\quad
\end{equation}
In these formulas we used the new parameters denoted by the same letters but
with subscript "$\times$". These new constants are defined by the relations:%
\begin{equation}
m_{\times}=M-m_{0},\quad b_{\times}=B-b_{0},\quad e_{\times}=Q_{e}+i Q_{m}-e_{0}
\label{20-3}%
\end{equation}%
\begin{equation}
a_{\times}+i\sigma_{s}=\frac{\Gamma\bar{\Gamma}\sqrt{c_{0}}}{\left(
a_{s}+i\sigma_{{s}}\right)  \sqrt{\mathcal{W}_{o}}}\text{ }, \label{20-4}%
\end{equation}%
\begin{equation}
\Gamma=\left(  m_{0}+ib_{0}\right)  \left(  m_{{s}}-ib_{{s}}\right)  -\overline{e}_{0} e_{s}  -\left(  a_{{s}}%
+i\sigma_{{s}}\right)  \left(  a_{0}-i\sigma_{{s}}-il\right),
\end{equation}
where parameters $c_{0,}M,B,Q_{e},Q_{m}$ have been defined by the
relations (\ref{c0}), (\ref{Physparameters}).

\paragraph{\emph{Conical singularities on the axis.}}
{If the conditions (\ref{20}) are satisfied, this does not mean yet that the
geometry on each part of the axis is regular since at the points of different
parts of the axis the local Euclidness of spatial geometry still may occur to
be violated. This behaviour of geometry on the sections $z=const$ looks like on the surface of a cone near its
vortex, where the ratio of the length of a circle (surrounding the vortex) to
its \textquotedblleft radius\textquotedblright\ is not equal to $2\pi$. In the
solution, on any surface $z=const$ intersecting the axis, the length and
radius of the circles $\rho=const$ are represented asymptotically for
$\rho\rightarrow0$ by the expressions $L=2\pi\sqrt{-g_{\varphi\varphi}}$ and
$R=\sqrt{f}\rho$ respectively. Besides that we note that the local Euclidness of the geometry on the parts of the axis of symmetry outside the sources, besides the condition $g_{t\varphi}\to 0$ implies for $\rho\to 0$ even more stronger condition  $g_{t\varphi}=O(\rho^2)$. Using this last condition and the mentioned above relation for metric coefficients  $g_{tt}g_{\varphi\varphi}-g_{t\varphi}^{2}=-\rho^{2}$, we obtain that for $\rho\rightarrow0$, the condition $L/(2\pi R)\rightarrow1$ is equivalent to the constraints:
\begin{equation}\label{PmPiPp}
P_{-}=P_{i}=P_{+}=1,\qquad P\equiv fg_{tt}\text{ },
\end{equation}
where $P_{-}$, $P_{i}$ and $P_{+}$ are the values of the product $fg_{tt}$
respectively on the negative, intermediate and positive parts of the axis. (In
accordance with the Einstein - Maxwell equations, the product $fg_{tt}$ is
constant on a part of the axis where }$g_{t\varphi}=0${, however, these
constants again may occur to be different for different disconnected parts of
the axis.) } To obtain the constrains on the parameters of our solution implied by (\ref{PmPiPp}), we have to analyze the behaviour of metric components on different parts of the axis of symmetry.
\medskip

\noindent$\underline{\text{\textit{Negative semi-infinite part of the axis:}
} \{\rho=0,\,-\infty<z<z_{1}-\sigma_{0}\}}$. On this part of the axis for bipolar coordinates we have the expressions
\begin{equation}
x_{1}=z_{1}-z,\quad x_{2}=l+z_{1}-z,\quad y_{1}=y_{2}=-1 \label{22}%
\end{equation}
and the metric component $g_{tt}$ and the conformal factor $f$ take the values \begin{equation}\label{gttmfm}
\begin{array}{l}
g_{tt} =\dfrac{[(z-z_{1})^{2}-\sigma_{0}^{2}][(z-z_{1}-l)^{2}-\sigma_{{s}%
}^{2}]}{c_{0}D_{-}\overline{D}_{-}},\\
\text{ }f =\dfrac{c_{0}D_{-}\overline{D}_{-}}{[(z-z_{1})^{2}-\sigma
_{0}^{2}][(z-z_{1}-l)^{2}-\sigma_{{s}}^{2}]},
\end{array}
\end{equation}
where $c_{0}$ has been defined in (\ref{c0}) and $D_{-}$ denotes the value of
$D$ (defined by (\ref{Dvalue})) on the negative semi-infinite part of the axis. These expressions show that on this part of the axis the corresponding condition from (\ref{PmPiPp}), i.e. $P_-=1$  is satisfied automatically.
\medskip

\noindent$\underline{\text{\textit{Positive semi-infinite part of the axis:}} \{\rho=0,\,z_{1}+l<z<\infty\}}$. On this part of axis the bipolar coordinates possess the expressions
\begin{equation}
x_{1}=z-z_{1},\quad x_{2}=z-z_{1}-l,\quad y_{1}=y_{2}=1 \label{24}%
\end{equation}
and here for the metric component $g_{tt}$ and for the conformal factor we have
\begin{equation}\label{gttpfp}
\begin{array}{l}
g_{tt} =\dfrac{[(z-z_{1})^{2}-\sigma_{0}^{2}][(z-z_{1}-l)^{2}-\sigma_{{s}
}^{2}]}{c_{0}D_{+}\overline{D}_{+}},\\
f =\dfrac{c_{0}D_{+}\overline{D}_{+}}{[(z-z_{1})^{2}-\sigma_{0}
^{2}][(z-z_{1}-l)^{2}-\sigma_{{s}}^{2}]}\text{ },
\end{array}
\end{equation}
where $D_{+}$ denotes the value of $D$ on the positive semi-infinite part of
the axis. From these expressions we see that the corresponding condition from (\ref{PmPiPp}) is satisfied identically, i.e. $P_+=1$ on the positive semi-infinite part of the axis.
\medskip

\noindent$\underline{\text{\textit{Intermediate part of the axis:}}
\{\rho=0,\,z_{1}+\sigma_{0}<z<z_{1}+l\}}$. At these points for bipolar coordinates we have the expressions
\begin{equation}
x_{1}=z-z_{1},\quad x_{2}=l+z_{1}-z,\quad y_{1}=1,\quad y_{2}=-1, \label{25}%
\end{equation}
and the corresponding expressions for $g_{tt}$ and the conformal factor $f$ are
\begin{align}
g_{tt}  &  =\dfrac{[(z-z_{1})^{2}-\sigma_{0}^{2}][(z-z_{1}-l)^{2}-\sigma_{{s}%
}^{2}]}{c_{0}D_{i}\overline{D}_{i}}\left(  \dfrac{1-2\delta}{1+2\delta
}\right)  ^{2},\text{ \ }\label{26}\\
f  &  =\dfrac{c_{0}D_{i}\overline{D}_{i}}{[(z-z_{1})^{2}-\sigma_{0}%
^{2}][(z-z_{1}-l)^{2}-\sigma_{{s}}^{2}]}\text{ },\nonumber
\end{align}
where $\delta$ is the same as defined already
by the expression (\ref{delta}). As it follows from these expressions, on the intermediate part of the axis the corresponding condition from (\ref{PmPiPp}), i.e. the equation $P_i=1$ is equivalent to the constraint
\begin{equation}\label{Zerodelta}
\delta=0.
\end{equation}

\paragraph{\emph{Magnetic and electric charges of the sources.}}
To obtain more realistic configurations, we have to exclude from the solution the total magnetic charge, i.e. to set $Q_m=0$ (see the expressions (\ref{Physparameters})) as well as the magnetic
charges of both sources. To calculate the physical values of magnetic charges
we should consider the magnetic
fluxes coming through closed space-like surfaces surrounding each charged center
and apply the Gauss theorem. In this way we can find the physical magnetic
charges $\mu_{0},\mu_{s}$ (as well as physical electric charges
$q_{0},q_{s}$ ) of each source calculating the corresponding
Komar-like integrals. The detailed procedure how to do this have been
described in the section "Physical parameters of the sources" in paper
\cite{AB2}. The results of these calculations are:
\begin{equation}\label{q0mu0}
q_{0}=\text{Re} (e_{0})+\operatorname{Re}F\text{ },\quad\mu_{0}=\operatorname{Im} (e_{0})+\operatorname{Im}F,
\end{equation}
\begin{equation}
q_{s}=\operatorname{Re}(e_{\times})-\operatorname{Re}F,\quad\mu_{s}
=\operatorname{Im}(e_{\times})-\operatorname{Im}F\text{ },
\end{equation}
where
\begin{equation}
F=e_{\times}\dfrac{(a_{{\times}}+i\sigma_{{s}%
}-im_{{\times}}+b_{{\times}})}{2(a_{{\times}}+i\sigma_{{s}})}-\dfrac
{\mathcal{L}_{0}\mathcal{H}_{0}}{2\mathcal{W}_{o}(a_{{\times}}+i\sigma_{{s}}%
)}\dfrac{(1+2\delta)}{(1-2\delta)}\text{ }, \label{30}%
\end{equation}
and we introduced here the new parameter polynomial:
\begin{equation}\label{L0}
  \mathcal{L}_{0} =\left[  (l+\sigma_{{s}})^{2}-\sigma_{0}^{2}\right]  e_{\times}
  +2\left[  X_{{\times}}-i(a_{{\times}}+i\sigma_{{s}})(l+\sigma_{{s}}%
-ia_{0})\right]  e_{0},
\end{equation}
where  $ X_{{\times}}$ was defined in (\ref{Xtimes}). The physically acceptable solution should satisfy
\begin{equation}\label{mu0mus}
\mu_{0}=\mu_{s}=0.
\end{equation}
These conditions mean that the physical magnetic charges of both sources vanish. It is useful to note here also that the total magnetic charge of the configuration $Q_m=\operatorname{Im} e_{0}+\operatorname{Im}e_{\times}=\mu_{0}+\mu_{s}$ and therefore, the conditions (\ref{mu0mus}) is sufficient for the condition $Q_m=0$ would be satisfied.

\paragraph{Summary for the equilibrium conditions.}
Here we present a list of the constraints on the parameters of our solution  providing a physical equilibrium of two interacting Kerr-Newman sources -- a Kerr-Newman black hole and a Kerr-Newman naked singulariry in their common gravitational and electromagnetic fields.
Choosing the multiplier $c_0$ in $f$ as it was determined in (\ref{c0}), we obtain an eleven-parameter solution which parameters for equilibrium configurations should satisfy the following five constraints (equilibrium conditions):
\begin{itemize}
\item[I.]{$\underline{\text{\textit{Absence of closed timelike curves on semi-infinite parts of the axis.}}}$ This \hfill\\ equilibrium conditions is equivalent to the vanishing of a NUT parameter $B$, and it implies in accordance with (\ref{Physparameters}) the constraint:
\begin{equation}
\text{Im}\left[  \dfrac{(m_{0}-ib_{0})(1+K_{1}+L_{1}+S_{0})+(m_{{s}}-ib_{{s}%
})(1+K_{2}+L_{2})}{1+S_{0}-(K_{1}+L_{1})(K_{2}+L_{2})}\right]  =0 \label{Eq1}%
\end{equation}}
\item[II.]{$\underline{\text{\textit{Absence of closed timelike curves on the axis between the sources.}}}$ From \hfill\\ (\ref{Omegai}), with the  condition $B=0$ taken into account, we obtain
\begin{equation}
\mathcal{W}_{o}\,\omega_{{\times}}\overline{\omega
}_{{\times}}=\mathcal{H}_{0}\overline{\mathcal{H}}_{0}\,\dfrac{(1+2\delta)}{(1-2\delta
)}
\end{equation}}
This simplifies  obviously, if we use here the constraint $\delta=0$ (see below)
\item[III.]{$\underline{\text{\textit{Absence of conical points on the axis of symmetry.}}}$
The choice (\ref{c0}) of the constant $c_0$ provides the absence of conical points on both semi-infinite parts of the axis. The cosntraint $\delta=0$ provides the absence of conical points on the part of the axis between the sources. This constraint, due to (\ref{delta}), can be presented in the form
\begin{equation}
m_{0}m_{{s}}+b_{0}b_{{s}}=\operatorname{Re}(e_{0}\overline{e}_{s})
\end{equation}}
\item[IV.]{$\underline{\text{\textit{Vanishing of magnetic charges of each of the sources.}}}$
To eliminate the physical magnetic charges $\mu_{0}, \mu_{s}$ of the sources, using (\ref{q0mu0})-(\ref{L0}) and  bearing in mind $\delta=0$,
we obtain the following two conditions:
\begin{equation}\label{magcharges}
\begin{array}{l}
\operatorname{Im}(e_{0})+\operatorname{Im}\left[e_{\times}
\dfrac{(a_{{\times}}+i\sigma_{{s}}-im_{{\times}}+b_{{\times}})}{2(a_{{\times}
}+i\sigma_{{s}})}-\dfrac{\mathcal{L}_{0}\mathcal{H}_{0}}{2\mathcal{W}_{o} (a_{{\times}}+i\sigma_{{s}})}\right] =0\\[2ex]
\operatorname{Im}(e_{\times})-\operatorname{Im}\left[e_{\times}
\dfrac{(a_{{\times}}+i\sigma_{{s}}-im_{{\times}}+b_{{\times}})}{2(a_{{\times}
}+i\sigma_{{s}})}-\dfrac{\mathcal{L}_{0}\mathcal{H}_{0}}{2\mathcal{W}_{o} (a_{{\times}}+i\sigma_{{s}})}\right]  =0
\end{array}
\end{equation}}
\end{itemize}
All these constraints possess rather complicated forms which do not allow to expect their simple and explicit solution in exact form. Because of that we consider the solution of these constraints  using a supplementary assumption that the distance $l$ separating the sources is rather large.

\section{Black~hole\,\,--\,\,naked~singularity~system~with\\ large~separating~distance}
The existence and properties of equilibrium configurations of two interacting Kerr - Newman sources can be studied effectively if we use some rather natural assumption that the $z$-distance $l$ which separates a black hole and a naked singularity is large enough in comparison with the linear sizes which characterize the sources. Namely, we assume that  the parameters (\ref{Parameters}) satisfy the condition:
\begin{equation}\label{Assumption}
d\equiv\max\{\vert m_{\scriptscriptstyle{0}}\vert,\,\vert a_{\scriptscriptstyle{0}}\vert,\, \vert b_{\scriptscriptstyle{0}}\vert,\,\vert e_{\scriptscriptstyle{0}}\vert, \vert m_{s}\vert,\,\vert a_{s}\vert,\,\vert b_{s}\vert,\,\vert e_{s}\vert\} \ll l
\end{equation}

\paragraph{\emph{On the absence in the solution (\ref{ErnstPotentials}) -- (\ref{factorf}) of other sources besides Kerr - Newman black hole and Kerr - Newman naked singularity.}}
We construct our solution generating a one soliton of the Kerr-Newman black hole backgound. By a construction, this solution (at list in the case of rather large z-distance separating the black hole and the location of the soliton (Kerr-Newman naked singularity) should not include other singularities which could be additional sources of gravitational and electromagnetic fields in this solution. To exhibit the absence of other curvature singularities (on the axis of symmetry or outside it) more clearly, we construct asymptotic representations of our solution in three \underline{\emph{overlapping}} space-time domains which, being considered together, cover the whole space-time region outside the two Kerr-Newman sources. In all three series expansions the same small parameter is used (see (\ref{Assumption})))
\[\varepsilon=\dfrac{d}{l}\ll 1
\]
and we consider a few first terms of the corresponding series expansions. However, we do not prove here the convergency of these $\varepsilon$-series because it could be rather tedious and it is not in the scope of our paper. However, it seems "highly likely" that these series expansions converge, because for large separating distance, one of these expansions represents a post-Newtonian approximations of different orders of the fields of one of the sources perturbed by the field of another source, while the other expansions describe small perturbations of the nearest zone of each of the sources by the external field of the other source.\footnote{It seems rather natural to expect that the leading terms of the asymptotic representations of our solution, constructed below,  for small enough $\varepsilon$ describe correctly the asymptotic behaviour of the solution in the corresponding overlapping space-time domains. However, if one has a sharp desire to prove the convergency of these series, we can mention some (presumably helpful) idea of this proof. Namely, our $\varepsilon$-series possess the forms $\sum a_n \varepsilon^n$, where each coefficient $a_n$ is a sum of terms of finite values which number grows with $n$. Therefore, for convergency of the series it is enough, for example, that the sums representing the coefficients $a_n$ would consist of the terms which number grows with $n$ not so fast as \underline{\emph{some}} fixed power $k$ of  $n$, i.e. if for all $n$ we have $a_n\lesssim n^k$ with some $k>0$. Then we can construct a majorant series which sum is the known Polylogarithm $(-k,\varepsilon)$ function. This function takes finite values for $0<\varepsilon<1$ and this proves a regular convergency of the corresponding series $\sum a_n \varepsilon^n$.}

To construct the asymptotic representations of our solution, we introduce two functions which are the "Eucledean" distances on the plane of Weyl coordinates $(\rho,z)$ from two points $(\rho=0,z=z_1)$ and $(\rho=0,z=z_2=z_1+l)$:
\[R_1=\sqrt{\rho^2+(z-z_1)^2},\qquad R_2=\sqrt{\rho^2+(z-z_1-l)^2}.
\]
The asymptotic representations of our solution we consider in three domains
\[(S0):\hskip2ex R_1\gg d\hskip1ex\text{and}\hskip1ex R_2\gg d,\qquad
(S1):\hskip2ex R_1\ll l,\qquad (S2):\hskip2ex R_2\ll l
\]
It is clear that pairs $(S_0, S_1)$ and $(S_0, S_2)$ of these domains overlap in the regions
\[\hbox{$(S0\cap S1):\hskip1ex  d\ll R_1\ll l$}\quad\qquad\text{and}\quad\qquad
\hbox{$(S0\cap S2):\hskip1ex  d\ll R_2\ll l$}
\]
and therefore, the regularity of asymptotic representations of our solution in the domains $S_0$, $S_1$ and $S_2$ will indicate the absence of any curvature singularities in the entire space-time region outside two Kerr-Newman sources which superposition of gravitational and electromagnetic  fields is described by our solution.

\noindent\paragraph{\emph{Asymptotic structure of the solution in the domain $S_0$}.} In this domain, i.e. at large distances from each of the sources (in comparison with  their typical "sizes" $\sim d$), we use an approximation in spirit of Bonnor \cite{Bonnor:2001}, in which we introduce (unlike Bonnor's paper), a small parameter $\varepsilon$ explicitly:
\begin{equation}\label{epsilon}
\left.\begin{array}{lclclcl}
m_0=\varepsilon \widehat{m}_0& a_0=\varepsilon \widehat{a}_0& b_0=\varepsilon \widehat{b}_0& e_0=\varepsilon \widehat{e}_0\\
m_s=\varepsilon \widehat{m}_s& a_s=\varepsilon \widehat{a}_s& b_s=\varepsilon \widehat{b}_s& e_s=\varepsilon \widehat{e}_s
\end{array}\quad\right\Vert\quad \varepsilon=\dfrac{d}{l} \ll 1,
\end{equation}
where, in accordance with our basic assumption (\ref{Assumption}), the parameter $d\ll l$. The parameter $l$, all new constant parameters with hats as well as the distances $R_1$ and $R_2$ also take some finite values.
In this domain, the Weyl coordinates $(\rho,z)$  can be expressed in terms of $R_1$ and $R_2$, and for bipolar coordinates we obtain
\[\begin{array}{lcl}
x_1=R_1+\dfrac{\rho^2\widehat{\sigma}_0^2}{2 R_1^3}\varepsilon^2+O(\varepsilon^4)&&
y_1=\dfrac{z-z_1}{R_1}-\dfrac{\rho^2(z-z_1)\widehat{\sigma}_0^2}{2 R_1^5}\varepsilon^2+O(\varepsilon^4)\\[2ex]
x_2=R_2+\dfrac{\rho^2\widehat{\sigma}_s^2}{2 R_2^3}\varepsilon^2+O(\varepsilon^4)&&
y_2=\dfrac{z-z_2}{R_2}-\dfrac{\rho^2(z-z_2)\widehat{\sigma}_s^2}{2 R_2^5}\varepsilon^2+O(\varepsilon^4)
\end{array}
\]
Substituting these expansions together with (\ref{epsilon}) into (\ref{ErnstPotentials})--(\ref{Ys}), we obtain the following asymptotic representation for the Ernst potentials of our solution:
\begin{equation}\begin{array}{l}\label{EFepsilon}
\mathcal{E}=1-2\left(\dfrac{\widehat{m}_0-i \widehat{b}_0}{R_1}+
\dfrac{\widehat{m}_s-i \widehat{b}_s}{R_2}\right)\varepsilon+\dfrac{\mathcal{A}(R_1,R_2)}{l (\widehat{a}_0^2+\widehat{\sigma}_0^2)(\widehat{a}_s+i\widehat{\sigma}_s) \widehat{\sigma}_s R_1^3 R_2^3}\, \varepsilon^2+O(\varepsilon^3),\\[3ex]
\Phi=\left(\dfrac{\widehat{e}_0}{R_1}+
\dfrac{\widehat{e}_s}{R_2}\right)\varepsilon+\dfrac{\mathcal{B}(R_1,R_2)}{l (\widehat{a}_0^2+\widehat{\sigma}_0^2)(\widehat{a}_s+i\widehat{\sigma}_s) \widehat{\sigma}_s R_1^3 R_2^3}\, \varepsilon^2+O(\varepsilon^3),
\end{array}
\end{equation}
where $\mathcal{A}(R_1,R_2)$ and $\mathcal{B}(R_1,R_2)$ are polynomials of the parameters with hats and of the distances $R_1$ and $R_2$. It is easy to check that coefficients in higher terms of these asymptotics possess the similar structures. The regular structures of the coefficients of asymptotic representations (\ref{EFepsilon}) indicates obviously the absence of any singularities of the Ernst potentials outside the $\varepsilon$-vicinities of the points $R_1=0$ and $R_2=0$, i.e. outside the small regions surrounding the sources -- a black hole and a naked singularity of Kerr - Newman types.

\paragraph{\emph{Asymptotic structure of the solution near the black hole (domain $S_1$)}} Near the black hole, we consider the domain with Weyl coordinates such that
\begin{equation}\label{asymp1}
R_1\ll l, \qquad\quad\Longrightarrow\quad\qquad
\rho\ll l\quad \text{and}\quad \vert z-z_1\vert \ll l
\end{equation}
In this domain, the values of $\rho$ and $z-z_1$ are of the order $\sim d$ where $d\ll l$ is a typical "size" of each of the sources (see (\ref{Assumption})). Therefore, we can introduce here, instead of (small enough) $\rho$ and $z-z_1$, the dimensionless coordinates with hats which take in this domain the finite velues:
\begin{equation}\label{rozhats}
\rho=\widehat{\rho}\, d\quad \text{and}\quad  z-z_1= \widehat{z}\, d
\end{equation}
Then, in accordance with (\ref{x2y2}), for bipolar coordinates $(x_2,y_2)$, we have
\begin{equation}\label{sx2sy2}
\left.\begin{array}{l}
  x_2=l \left[1-\widehat{z}\varepsilon+\dfrac{1}{2}\widehat{\rho}{}^2\varepsilon^2+ \dfrac{1}{2}\widehat{\rho}{}^2 \widehat{z}\varepsilon^3+O(\varepsilon^4)\right],\\
  y_2=-1+\dfrac{1}{2}\widehat{\rho}{}^2\varepsilon^2+ \widehat{\rho}{}^2 \widehat{z}\,\varepsilon^3+O(\varepsilon^4),
\end{array}\quad\right\Vert\quad \varepsilon=\dfrac{d}{l}\ll 1.
\end{equation}
It is convenient to return now to the Weyl coordinates using (\ref{rozhats}) and express these coordinates in accordance with (\ref{Weyl}) in terms of polar coordinates $(x_1,y_1)$ most adapted for the first source. Then we obtain from (\ref{sx2sy2}) the series
\begin{equation}\label{x2y2-x1y1}
\begin{array}{l}
  x_2=l\left[-\dfrac{x_1 y_1}{d}\,\varepsilon+\dfrac{(x_1^2-\sigma_0^2)(1-y_1^2)}{2 d^2}\varepsilon^2+\dfrac{x_1 y_1(x_1^2-\sigma_0^2)(1-y_1^2)}{2 d^3}\varepsilon^3+O(\varepsilon^4)\right] \\[2ex]
  y_2=-1+\dfrac{(x_1^2-\sigma_0^2)(1-y_1^2)}{2 d^2}\,\varepsilon^2+\dfrac{x_1 y_1 (x_1^2-\sigma_0^2)(1-y_1^2)}{d^3}\,\varepsilon^3+O(\varepsilon^4)
\end{array}
\end{equation}
Suibstituting these series into the Ernst potentials (\ref{ErnstPotentials}) of our solution and taking into account the expressions (\ref{CalR1R2})--(\ref{Ys}), we obtain the following expansions
\begin{equation}\label{Source1}
\begin{array}{l}
\mathcal{E}=\dfrac{x_1-m_0+i b_0+i a_0 y_1}{x_1+m_0-i b_0+i a_0 y_1}+
\dfrac{\mathcal{E}_1(x_1,y_1)}{(a_s+i\sigma_s)(x_1+m_0-i b_0+i a_0 y_1)^2 d} \,\varepsilon+O(\varepsilon^2)\\[2ex]
\Phi=\dfrac{e_0}{x_1+m_0-i b_0+i a_0 y_1}+
\dfrac{\mathcal{F}_1(x_1,y_1)}{(a_s+i\sigma_s)(x_1+m_0-i b_0+i a_0 y_1)^2 d}\,\varepsilon +O(\varepsilon^2)
\end{array}
\end{equation}
where $\mathcal{E}_1(x_1,y_1)$ and $\mathcal{F}_1(x_1,y_1)$ are quadratic polynomials of coordinates $x_1$ and $y_1$ with coefficients which are polynomial of parameters of the solution:
\begin{equation}\label{E1F1}
\begin{array}{l}
\mathcal{E}_1(x_1,y_1)=2 i \overline{e}_s[e_0(m_s-i b_s)-e_s(m_0-i b_0)](x_1+i a_0 y_1)\\
\phantom{\mathcal{E}_1(x_1,y_1)}+2(m_0-i b_0)(a_s+i\sigma_s)[(m_0-i b_0)(m_s+i b_s)-e_0 \overline{e}_s ]\\
\phantom{\mathcal{E}_1(x_1,y_1)}-2(a_s+i\sigma_s)(x_1+i a_0 y_1)[(m_s-i b_s)(x_1+i a_0 y_1)+2 i a_s(m_0-i b_0)]\\[1ex]
\mathcal{F}_1(x_1,y_1)=e_s(a_s+i\sigma_s)(x_1+m_0-i b_0+i a_0 y_1)^2\\
\phantom{\mathcal{F}_1(x_1,y_1)}+i e_s[(m_0-i b_0)(m_s+i b_s)-e_0 \overline{e}_s](x_1+m_0-i b_0+i a_0 y_1)\\
\phantom{\mathcal{F}_1(x_1,y_1)}-2(a_s+i\sigma_s)(m_0-i b_0)[e_s(x_1+m_0-i b_0+i a_0 y_1)+i e_0(a_s+b_s)]\\
\phantom{\mathcal{F}_1(x_1,y_1)}+[e_0(m_s-i b_s-i a_s+\sigma_s)-e_s(m_0-i b_0)]\\
\phantom{\mathcal{F}_1(x_1,y_1)}\times [i e_0 \overline{e}_s-(a_s+i \sigma_s)(x_1+m_0-i b_0+i a_0 y_1)]
\end{array}
\end{equation}

In each of the series (\ref{Source1}) the first terms ($\sim \varepsilon^0$) correspond exactly to the Ernst potentials of the Kerr-Newman black hole ($\sigma_0^2>0$), while the next terms ($\sim \varepsilon,\varepsilon^2,\ldots$) describe small perturbations of this black hole by a distant Kerr-Newman naked singularity ($\sigma_s^2<0$). The equations (\ref{Source1}), (\ref{E1F1}) show that this perturbation of the Ernst potentials near the black hole is completely regular.

\medskip
\paragraph{\emph{Asymptotic structure of the solution near the naked singularity   (domain $S_2$).}}   Similarly to the consideration given just above of the region
near the black hole, we consider now the region near the naked singularity (the source 2) which can be described in the Weyl coordinates by the conditions
\begin{equation}\label{asymp2}
R_2\ll l, \qquad\quad\Longrightarrow\quad\qquad
\rho\ll l\quad \text{and}\quad \vert z-z_2\vert \ll l
\end{equation}
In this domain, in accordance with (\ref{x1y1}), for bipolar coordinates, $(x_1,y_1)$ we have
\begin{equation}\label{sx1sy1}
\begin{array}{l}
  x_1= l\left[1+\dfrac{z-z_2}{d}\varepsilon+\dfrac{\rho^2}{2 d^2}\varepsilon^2-\dfrac{(z-z_2)\rho^2}{2 d^3}\varepsilon^3+O(\varepsilon^4)\right] \\[1ex]
  y_1=1-\dfrac{\rho^2}{2 d^2}\,\varepsilon^2+\dfrac{(z-z_2)\rho^2}{d^3}\,\varepsilon^3+O(\varepsilon^4)
\end{array}
\end{equation}
Using in (\ref{sx1sy1}) the expressions (\ref{Weyl}), we obtain
\begin{equation}\label{x1y1-x2y2}
\begin{array}{l}
x_1=l\left[1+\dfrac{x_2 y_2}{d}\varepsilon+\dfrac{(x_2^2-\sigma_s^2)(1-y_2^2)}{2 d^2}\,\varepsilon^2-\dfrac{x_2 y_2(x_2^2-\sigma_s^2)(1-y_2^2)}{2 d^3}\,\varepsilon^3+O(\varepsilon^4)\right] \\[2ex]
 y_1=1-\dfrac{(x_2^2-\sigma_s^2)(1-y_2^2)}{2 d^2}\,\varepsilon^2+\dfrac{x_2 y_2 (x_2^2-\sigma_s^2)(1-y_2^2)}{d^3}\,\varepsilon^3+O(\varepsilon^4)
\end{array}
\end{equation}
Substituting these series into (\ref{CalR1R2})--(\ref{Ys}), we obtain the following asymptotics for the Ernst potentials of our solution in the neighbourhood of the source 2:
\begin{equation}\label{Source2}
\begin{array}{l}
\mathcal{E}=\dfrac{x_2-m_s+i b_s+i a_s y_2}{x_2+m_s-i b_s+i a_s y_2}+
\dfrac{\mathcal{E}_2(x_2,y_2)}{k_0 (x_2+m_s-i b_s+i a_s y_2)^2 d}\,\varepsilon+O(\varepsilon^2)\\[2ex]
\Phi=\dfrac{e_s}{x_2+m_s-i b_s+i a_s y_2}+
\dfrac{\mathcal{F}_2(x_2,y_2)}{k_0 (x_2+m_s-i b_s+i a_s y_2)^2 d}\,\varepsilon+O(\varepsilon^2)
\end{array}
\end{equation}
where $k_0=\sigma_s (a_s+i\sigma_s)(a_0^2+\sigma_0^2)$ and the coefficients
$\mathcal{E}_2(x_2,y_2)$ and $\mathcal{F}_2(x_2,y_2)$ are quadratic polynomials of coordinates $x_2$ and $y_2$ with coefficients which are polynomials of parameters of the solution $m_0$,$m_s$,$a_0$,$a_s$,$b_0$,$b_s$,$e_0$,$e_s$ and $\sigma_0$,$\sigma_s$.

In the series (\ref{Source2}) the first terms ($\sim \varepsilon^0$) correspond to a Kerr-Newman naked singularity ($\sigma_s^2<0$), while the subsequent terms ($\sim \varepsilon, \varepsilon^2,\ldots$) describe small perturbations of the gravitational and electromagnetic fields of this naked singularity by a distant Kerr-Newman black hole.
The explicit expressions for $\mathcal{E}_2(x_2,y_2)$ and $\mathcal{F}_2(x_2,y_2)$ are very similar to the expressions for the coefficients (\ref{E1F1}), but these are more longer and we do not present them here. However, it is important that like the expressions (\ref{E1F1}), these expressions show that the rather distant Kerr-Newman black hole (Source~1) produces a regular perturbation in the vicinity of the "critical surface" (i.e. in the region $x_2 \ge 0$) surrounding the Kerr-Newman naked singularity (Source~2).

Thus, in the above we had shown that, at least for rather large values of the coordinate $z$-distance $l$ separating the black hole and the naked singularity, the asymptotical behaviour of the Ernst potentials is regular everywhere outside the black hole horizon ($x_1\ge \sigma_0$) and "critical surface" of the naked singularity ($x_2\ge 0$) and therefore, there are no any additional singularity can be expected in this solution besides two Kerr-Newman sources mentioned above.

\paragraph{\emph{Solution of equilibrium conditions (\ref{Eq1})-(\ref{magcharges})}}
In this section, we describe an asymptotic construction of solution of equilibrium conditions for our black hole - naked singularity system described by the solution (\ref{ErnstPotentials}) - (\ref{factorf}) of Einstein - Maxwell equations. Our construction is based on the main assumption (\ref{Assumption}) but we use now more detail asymptotic representation of the parameters of the solution as power series in terms of a small parameter $\varepsilon \ll 1$:
\begin{equation}\label{epsseries}
\begin{array}{ll}
m_0=m_{0\cdot}\varepsilon+m_{0\cdot\cdot}\varepsilon^2+ m_{0\cdot\cdot\cdot}\varepsilon^3+O(\varepsilon^4),&\hskip-1ex
m_s=m_{s\cdot}\varepsilon+m_{s\cdot\cdot}\varepsilon^2+ m_{s\cdot\cdot\cdot}\varepsilon^3+O(\varepsilon^4)\\[1ex]
b_0=b_{0\cdot}\varepsilon+b_{0\cdot\cdot}\varepsilon^2+ b_{0\cdot\cdot\cdot}\varepsilon^3+O(\varepsilon^4),&\hskip-1ex
b_s=m_{s\cdot}\varepsilon+b_{s\cdot\cdot}\varepsilon^2+ b_{s\cdot\cdot\cdot}\varepsilon^3+O(\varepsilon^4)\\[1ex]
a_0=a_{0\cdot}\varepsilon+a_{0\cdot\cdot}\varepsilon^2+ a_{0\cdot\cdot\cdot}\varepsilon^3+O(\varepsilon^4),&\hskip-1ex
a_s=a_{s\cdot}\varepsilon+a_{s\cdot\cdot}\varepsilon^2+ a_{s\cdot\cdot\cdot}\varepsilon^3+O(\varepsilon^4)\\[1ex]
e_0=e_{0\cdot}\varepsilon+e_{0\cdot\cdot}\varepsilon^2+ e_{0\cdot\cdot\cdot}\varepsilon^3+O(\varepsilon^4),&\hskip-1ex
e_s=e_{s\cdot}\varepsilon+e_{s\cdot\cdot}\varepsilon^2+ e_{s\cdot\cdot\cdot}\varepsilon^3+O(\varepsilon^4)
\end{array}
\end{equation}
where the number of dots in the suffices of coefficients characterizes the order of approximation. The z-distance $l$ is considered as given parameter of some finite value. Substitution of these expansions into the equilibrium conditions (\ref{Eq1})-(\ref{magcharges}) leads to the following leading order equations:
\[\begin{array}{lcl}
\text{(I)}\quad b_{0\cdot}+b_{s\cdot}=0,&&\text{(II)}\quad b_{s\cdot}=0\\[1ex]
\text{(III)}\quad
m_{0\cdot}m_{s\cdot}+b_{0\cdot}b_{s\cdot}-\text{Re}(e_{0\cdot} \overline{e}_{s\cdot})=0&&
\text{(IV)}\quad \text{Im}(e_{0\cdot})= \text{Im}(e_{s\cdot})=0
\end{array}
\]
It is easy to see from these leading order equations
that in this approximation we obtain the equilibrium configurations which depend on seven real parameters $m_{0\cdot}$, $a_{0\cdot}$, $e_{0\cdot}$,\, $m_{s\cdot}$, $a_{s\cdot}$, $e_{s\cdot}$ and $l$, which should satisfy the only relation
\[m_{0\cdot}m_{s\cdot}-e_{0\cdot} e_{s\cdot}=0.
\]
To conclude our present consideration, we note that in the leading approximation, the parameters mentioned just above represent respectively the physical parameters of the system -- masses, angular momentums per unit masses  and electric charges of the sources respectively. In this case, the total mass, electric charge and angular momentum of equilibrium configuration are
\[M_{tot\cdot}= m_{0\cdot}+m_{s\cdot},\quad
Q_{tot\cdot}= e_{0\cdot}+e_{s\cdot},\quad
J_{tot\cdot}=m_{0\cdot}a_{0\cdot}+m_{s\cdot}a_{s\cdot}
\]
However, in the subsequent approximations of the field of interacting sources these parameters loose their "individual" physical interpretations. (Just this phenomenon was observed (in the exact form) in our paper \cite{AB3} where the static limit of the field of two Kerr - Newman sources was considered.)

The equations of the subsequent orders which arise from the equilibrium conditions (\ref{Eq1})-(\ref{magcharges}), allow to calculate subsequently the other coefficients of the expansions (\ref{epsseries}). This leads to more precise description of equilibrium configurations of two Kerr - Newman sources under consideration and of influence of their interaction on the physical paremeters of each of these sources.
However, the detail analysis of higher orders of the used approximation is not in the scope of the present paper, in which our purpose was to present the corresponding exact solution and to show the existence of six-parametric family of equilibrium configurations.
More detail description of this family of equilibrium configurations as well as consideration of superposition of fields of other types of Kerr-Newman sources we are going to postpone for a future work.

\subsection*{Acknowledgements}
The authors thank the Referee and the Editor respectively for useful questions and comments, which urged us to clarify some points and to include into the manuscript more detail description of the structure of our solution.

The work of GAA was supported in parts by the Russian Foundation for Basic Research (grant 18-01-00273 a).

VAB would like to express his gratitude to the Max-Planck-Institute for Gravitational Physics(Albert Einstein Institute) at Golm (Germany) for the fruitful collaboration, hospitality and financial support.


\bigskip

\begin{thebibliography}{99}                                                                                               %

\bibitem {Bon}W. B. Bonnor "The equilibrium of a charged test particle in the
field of a spherical charged mass in general relativity", Class. Quant. Grav.,
\textbf{10}, 2077 (1993).

\bibitem {Per}G. P. Perry and F. I. Cooperstock "Electrostatic equilibrium of
two spherical charged masses in general relativity", Class. Quant. Grav.,
\textbf{14}, 1329 (1997); arXiv:gr-qc/9611066.

\bibitem {Alekseev-Belinski:2007} G. A. Alekseev and V. A. Belinski "Equilibrium configurations of two charged masses in General Relativity", Phys.Rev\textit{.} \textbf{D76,}
021501(R) (2007); arXiv:gr-qc/0706.1981.

\bibitem {AB2}G. A. Alekseev and V. A. Belinski "Superposition of Fields of
Two Reissner-Nordstrom Sources", in Proceedings of the Eleventh Marcel
Grossmann Meeting on General Relativity (Berlin, Germany 2007), Part A:
Plenary and Review talks, eds. H. Kleinert, R.T. Jantzen and R. Ruffini (World
Scientific, Singapore, 2008), p. 543; arXiv:gr-qc/0710.2515.

\bibitem {BZ1}V. Belinski and V. Zakharov "Integration of the Einstein
Equations by means of the inverse scattering problem technique and
construction of exact soliton solutions", Sov. Phys. JETP, \textbf{48, }985 (1978).

\bibitem {BZ2}V. Belinski and V. Zakharov "Stationary gravitational solitons
with axial symmetry", Sov. Phys. JETP \textbf{50, }1 (1979).

\bibitem {A1}G. A. Alekseev "N-soliton solutions of Einstein-Maxwell
equations", Pis'ma Zh. Eksp. Teor. Fiz. \textbf{32}, 301 (1980); English
transl. JETP Lett. \textbf{32, }277 (1981).

\bibitem {A2}G. A. Alekseev "Exact Solutions in the General Theory of
Relativity" \textit{Trudy Metem. Inst. Steklova}, \textbf{176, }211 (1987);
English transl. \textit{Proceedings of the Steklov Institute of Mathematics}
(American Mathematical Society, Providence, Rhode Island), issue 3 of 4 p. 215 (1988).

\bibitem{BV} V.Belinski and E.Verdaguer "Gravitational Solitons", Cambridge
University Press (2001).

\bibitem {AB3}G. Alekseev and V. Belinski \textquotedblleft Soliton Nature of
Equilibrium State of Two Charged Masses in General
Relativity\textquotedblright, International Journal of Modern Physics, Conference Series, vol.12, p.10 (2012); arXiv:1103.0582 [gr-qc].

\bibitem {A3}G. A. Alekseev "Twelve-parametric electrovacuum two-soliton
solution - the external field of two interacting Kerr-Newman sources", in
\textit{Abstracts of contributed papers 11th International Conference on
General Relativity and Gravitation}, Vol. \textbf{1} (Stockholm, Sweeden,
1986), p. 227.

\bibitem{Bonnor:2001} W.B. Bonnor, "The interactions of charged, spinning, magnetized masses",  Class. Quantum Grav. {\bf 18}, 2853-2863 (2001)

\end{thebibliography}
\end{document}